\documentclass[
  reprint,
  superscriptaddress, 
  amsmath,amssymb,
  aps,
  prl
]{revtex4-2}

\usepackage{graphicx}
\usepackage{dcolumn}
\usepackage{bm}
\usepackage{dsfont}
\usepackage{tikz}
\usepackage{indentfirst}
\usepackage{multirow}

\usepackage[colorlinks,citecolor=blue,linkcolor=red,urlcolor=magenta]{hyperref}

\newcommand{\supplementarysection}{%
  \setcounter{figure}{0}
  \let\oldthefigure\thefigure
  \renewcommand{\thefigure}{S\oldthefigure}
  \setcounter{section}{0}
  \let\oldthesection\thesection
  \renewcommand{\thesection}{S\oldthesection}
  \setcounter{equation}{0}
  \let\oldtheequation\theequation
  \renewcommand{\theequation}{S\oldtheequation}
  \setcounter{table}{0}
  \let\oldthetable\thetable
  \renewcommand{\thetable}{S\oldthetable}
}
\newcommand{\remove}[1]{}

\newcommand{\bi}{\begin{itemize}}
\newcommand{\ei}{\end{itemize}}
\newcommand{\be}{\begin{enumerate}}
\newcommand{\ee}{\end{enumerate}}

\newenvironment{dfn}{{\vspace*{1ex} \noindent \bf Definition }}{\vspace*{1ex}}

\newcommand{\nn}{\nonumber}  %

\newcommand{\Eq}[1]{Eq.~(\ref{#1})}

	\newcommand{\beq}{\begin{eqnarray}}
	\newcommand{\eeq}{\end{eqnarray}}
	\newcommand{\bea}{\begin{eqnarray}\begin{aligned}}
	\newcommand{\eea}{\end{aligned}\end{eqnarray}}

\definecolor{forestgreen}{RGB}{34,139,34}

\begin{document}
\title{Half dualization and non-invertible particle--vortex duality defect on lattice}
\author{Zhi-Qiang Gao}
\email{zqgao@berkeley.edu}
\affiliation{Department of Physics, University of California, Berkeley, California 94720, USA}

\begin{abstract}
We introduce half dualization as a general principle to construct non-invertible duality defects. The construction performs a duality transformation only in a subregion of spacetime, leaving an interface between the original theory and its dual. A non-invertible duality defect can be hosted on this interface. Half dualization is applicable whenever a local duality transformation is available, and does not depend on the spacetime dimension. We demonstrate the construction procedure in the (2+1)-dimensional Villainized charge-$n$ XY model on a cubic lattice. Half dualization yields a non-invertible particle--vortex duality defect whose fusion with its orientation reverse produces a (1+1)-dimensional $\mathbb{Z}_n$ gauge theory on the fusion surface. We then apply half dualization to the $\mathbb{Z}_n$-gauged XY model relevant to the 3D XY$^\ast$ transition. In the gauged model, the half dualization interface hosts a gauge covariant duality wall, rather than a genuine gauge invariant duality defect. It flows to an invertible duality defect in infrared when the $\mathbb{Z}_n$ gauge theory is confined or Higgsed.
\end{abstract}


\maketitle

{\it Introduction.---}
Symmetry is a central organizing principle in modern theoretical physics. Recently, the symmetry principle has been generalized far beyond group actions on local operators. Higher form symmetries act on extended operators, while non-invertible symmetries provide a further extension in which the symmetry operations need not be group-like. In the defect formulation, an ordinary symmetry is represented by a topological codimension-one operator whose fusion realizes the group law, and fusing a defect with its orientation reverse gives the identity. Non-invertible symmetries keep the topological nature of the defect, but their fusion rules are no longer group-like. A hallmark is that fusing a defect with its orientation reverse does not give only the identity~\cite{Shao2024}. Non-invertible symmetries appear in a wide range of quantum field theories. In (1+1)D, they are naturally realized by Verlinde lines in conformal field theories (CFTs)~\cite{Petkova2001,Chang2021,Thorngren2021,Sharpe2022}. In higher dimensions, TQFTs and gauge theories provide many further examples, including non-Abelian Wilson lines, Gukov--Witten defects, topological surface operators in (2+1)D TQFTs, and defects in (3+1)D gauge theories~\cite{Gukov2007,Gukov2008,Wen2019NI,Rudelius2020,Nguyen2021,Heidenreich2021,Kaidi2022,Antinucci2022,You2022,Cdova2022,Arias2023,Yu2023,Benini2023,McGreevy2023,Kapustin2010,Fuchs2013,Lan2015,Carqueville2018,Koide2021,Choi2022,Zheng2022,Choi2022a,Choi2023}.

Beyond these examples, two systematic routes to construct non-invertible defects are higher gauging and half gauging. In higher gauging, gauging a discrete higher form symmetry along a higher codimension submanifold leaves the bulk theory unchanged and produces a topological defect~\cite{Roumpedakis2023}. In half gauging, gauging a discrete symmetry only in half of spacetime produces an interface between the original theory and the gauged theory~\cite{Choi2022a}. When the two theories are equivalent, this interface becomes a self-duality defect, often non-invertible. However, the simplest self-dual half-gauging realization is most efficient in even spacetime dimension. Gauging a $p$-form discrete symmetry in $D$ dimensions typically produces a dual $(D-p-2)$-form symmetry, and thus self-duality imposes $p=D-p-2$, {\it i.e.}, $D=2(p+1)$ being even~\cite{Shao2024}. In (2+1)D, half gauging can still be implemented, but generally requires a more elaborate setup, such as gauging paired $0$-form and $1$-form symmetries~\cite{Cui2024}.

The success of half gauging suggests a broader lesson: applying an equivalence only to part of spacetime can lead to non-invertible defects. In (2+1)D, such equivalences are abundant because {\it duality} is ubiquitous. The paradigmatic example is bosonic particle--vortex duality between the XY model and the Abelian Higgs model~\cite{Villain_1975,Banks1977,Peskin1978,Dasgupta:1981aa,Fisher:1989aa}. Boson-fermion duality gives another important example and underlies a web of (2+1)D dualities~\cite{Aharony2016,Seiberg2016,Karch2016,Son2015,Kachru2015,Mross2016,Metlitski2016,Hsin2016,Aharony2017,Wang2017,Max2017,Huang:2021,Han2022,Han2023,Huang:2023}, where some of these dualities have exact lattice realizations~\cite{Chen2018,Aitken2018,Son2019}. Discovery of deconfined quantum criticality also inspires a large class of (2+1)D dualities~\cite{Senthil:2004aa,Senthil:2004ab,Motrunich:2004,Grover2007,Grover2008,Wangfa2015,Roberts2019,Senthil2019,Zhang2023,Lu2023,Wang2024,Pace2024,Wu2024,Gao2026dqcp}.

In this work, we introduce half dualization as a general principle to construct duality defects, including non-invertible ones. Starting from a lattice model, half dualization performs a local duality transformation on a subregion of the lattice, creating an interface between the original model and the dual model. Such an interface is topological in the sense that local changes of the cut are implemented by local duality transformations and leave the partition function invariant. Unlike half gauging, half dualization is not tied to a specific spacetime dimension. It applies whenever an exact lattice duality transformation is available. We illustrate the procedure using a (2+1)-dimensional charge-$n$ XY model~\cite{Lee:1985aa,Gao2022}, and show the existence of a non-invertible particle--vortex duality defect for $n>1$. This naturally generalizes to higher dimensions. We further apply half dualization to the $\mathbb{Z}_n$-gauged XY model. We find that the non-invertible duality defect becomes a gauge covariant duality wall, instead of a genuine duality defect. In IR, it flows to an invertible duality defect when the $\mathbb{Z}_n$ gauge field is Higgsed or confined.

{\it Half dualization.---}
Consider a lattice theory $\mathcal T[\phi]$ on a $D$-dimensional Euclidean spacetime lattice $L$ with local interactions~\footnote{Here by local interactions we mean there exists a cutoff $\ell$ in distance, beyond which microscopic variables are not coupled.}. The microscopic variables $\phi$ may live on sites, links, plaquettes, or more general cells of $L$. Let a codimension-one cut divide the lattice into two regions $L_1$ and $L_2$, and assume that $\mathcal{T}$ admits a lattice duality transformation to a dual theory $\widetilde{\mathcal{T}}[\tilde{\phi}]$. Half dualization performs this duality transformation only in $L_2$. It produces an interface coupling between the original variables $\phi$ on the boundary $\Sigma\subset L_1$ to the dual variables $\tilde{\phi}$ on the dual boundary $\widetilde{\Sigma}$. At partition function level, 
\beq
Z=\sum_{\phi,\tilde{\phi}}W_1[\phi]\times\mathcal{D}[\phi,\tilde{\phi}]\times\widetilde{W}_2[\tilde{\phi}] ,
\eeq
where $\mathcal{D}$ represents the duality wall. On the lattice it is supported in a finite-thickness region $\Sigma\bigcup\widetilde{\Sigma}$, and in the continuum limit it becomes a codimension-one surface. When $\mathcal{D}$ is gauge invariant as a standalone object, it defines a genuine duality defect between $\mathcal{T}$ and $\widetilde{\mathcal{T}}$. Its action on operators is determined by moving operators across, $O[\phi]\,\mathcal{D}=\mathcal{D}\,\tilde{O}[\tilde{\phi}]$. The invertibility of the duality defect can be derived straightforwardly from $\mathcal{T}$ and $\widetilde{\mathcal{T}}$.

{\it Charge-$n$ XY model and half duality transformation.---}
To illustrate the idea of half dualization, we consider the classical charge-$n$ XY model defined on a 3D cubic lattice~\cite{Lee:1985aa,Gao2022},
\beq
\beta H=\sum_{\left<\mathbf{ij}\right>}-K\cos{(n\theta_\mathbf{i}-n\theta_\mathbf{j})}-K^\prime\cos{(\theta_\mathbf{i}-\theta_\mathbf{j})},\label{eq:H}
\eeq
where $\theta_\mathbf{i}\in[0,2\pi)$ is a classical $U(1)$ variable defined on site $\mathbf{i}$, $n$ is a positive integer, $K,K^\prime>0$ are the coupling constants, and $\beta$ is the inverse temperature. Physically, charge-$n$ XY model \Eq{eq:H} describes the Josephson coupling of a vestigial charge-$2ne$ superconductor~\cite{Berg2009}. Therefore, $\theta_\mathbf{i}$ describes the phase of a microscopic charge-$2e$ Cooper pair, while $e^{in\theta_\mathbf{i}}$ is the charge-$2ne$ superconducting order parameter. Therefore, both $e^{i\theta_\mathbf{i}}$ and $e^{in\theta_\mathbf{i}}$ are physical operators. When $K^\prime<n^2K$, Hamiltonian \Eq{eq:H} on each link possesses local minima near $\theta_\mathbf{i}-\theta_\mathbf{j}=2\pi p/n$ with $p\in\mathbb{Z}_n$, which is crucial for the onset of non-invertible defect. For clarity, it suffices to consider $K^\prime=0$ in the following derivation of half dualization. The effect of the $K^\prime$ term is addressed in the End Matter. With $K^\prime=0$, the charge-$n$ XY model has two phases, the boson gapped phase for small $K$ and the superfluid (SF) phase for large $K$, separated by a 3D XY transition. However, it is not equivalent to an ordinary charge-$1$ XY model, as the physical operator content is different. Under Villain approximation~\cite{Villain_1975}, the partition function of the charge-$n$ XY model with $K^\prime=0$ is,
\beq
Z=\int D[\theta]\sum_{\{m_\mathbf{ij}\}}\exp{\bigg[-\frac{K}{2}\sum_{\left<\mathbf{ij}\right>}(n\theta_\mathbf{i}-n\theta_\mathbf{j}-2\pi m_\mathbf{ij})^2\bigg]},\label{eq:Z0}
\eeq
where integer variable $m_\mathbf{ij}$ is introduced on each link. In what follows, we treat \Eq{eq:Z0} as the starting point of the duality transformation. 

\begin{figure}[htbp]
  \centering
  \includegraphics[width=0.55\linewidth]{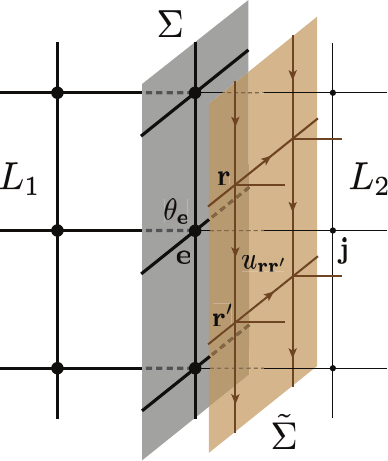}
  \caption{Duality defect constructed from half dualization of the charge-$n$ XY model on cubic lattice. The duality transformation is performed only on sites and links in $L_2$, which yields an integer variable $u_\mathbf{rr^\prime}$ defined on dual lattice links $\left<\mathbf{rr^\prime}\right>$ marked in brown. The duality defect is supported on the union of the boundary $\Sigma$ and the dual boundary $\tilde{\Sigma}$, colored in grey and brown, respectively. It is composed by the coupling between $\theta_\mathbf{e}$ on $\Sigma$ and curl of $u_\mathbf{rr^\prime}$ on $\tilde{\Sigma}$.}
  \label{fig:dual}
\end{figure}

Now consider a 2D wall $\Sigma$ which separates the 3D lattice $L$ into two sides, $L_1$ and $L_2$, such that $L=L_1\bigcup L_2$. Without loss of generality, we set $\Sigma\subset L_1$. The half duality transformation is performed in parallel to Ref.~\cite{Gao2022}, which yields (See End Matter for detailed derivation),
\beq
Z=\int D[\theta]\prod_{\left<\mathbf{ij}\right>\in L_1}\sum_{m_\mathbf{ij}}\exp\bigg[-\frac{K}{2}(n\theta_\mathbf{i}-n\theta_\mathbf{j}-2\pi m_\mathbf{ij})^2\bigg]\nn\\
\times\!\!\prod_{\left<\mathbf{ij}\right>\in L_2}\sum_{u_\mathbf{rr^\prime}}\exp\bigg[-\frac{(\nabla\times {\bm u})^2_\mathbf{ij}}{2K}\bigg]\exp{\sum_{\mathbf{e}\in\Sigma}in\theta_\mathbf{e}(\nabla\times {\bm u})_\mathbf{e}}.\nn
\eeq
Here $u_\mathbf{rr^\prime}$ is a dual integer field defined on the dual lattice links in $L_2$. Its curl appearing in the kinetic energy term is defined on the dual lattice plaquette that pierces the lattice link $\left<\mathbf{ij}\right>$. In particular, $(\nabla\times {\bm u})_\mathbf{e}$ pierces a lattice link ended at $\mathbf{e}\in\Sigma$. 
The partition function is clearly factorized into three parts
\beq
Z=\int D[\theta]\sum_{\{{\bm u}\}}W_1[\theta]\times\widetilde{W}_2[{\bm u}]\times \mathcal{D}[\theta,{\bm u}].\label{eq:Zf}
\eeq
Here $W_1$ is the Boltzmann weight of the Villainized charge-$n$ XY model \Eq{eq:Z0} defined on $L_1$. $\widetilde{W}_2$ represents the particle--vortex dual theory of the charge-$n$ XY model defined on the dual lattice of $L_2$,
\beq
\widetilde{W}_2[{\bm u}]=\exp\sum_\square-\frac{(\nabla\times {\bm u})^2}{2K},\label{eq:Z2}
\eeq
It describes a lattice Abelian Higgs model, upon introducing an integer vortex current $J_\mathbf{rr^\prime}$ and converting $u_\mathbf{rr^\prime}$ to a noncompact~\footnote{Here by noncompact we mean the gauge potential takes the value of $\mathbb{R}$, rather than over a compact interval. However, as it couples to an integer-valued vortex current, the allowed monopole insertions carry integer units of $2\pi$ flux according to Dirac's quantization condition.} $U(1)$ gauge field $a_\mathbf{rr^\prime}$ by Poisson's summation~\cite{Lee:1985aa,Gao2022}. The last piece in \Eq{eq:Zf} is the duality defect $\mathcal{D}$,
\beq
\mathcal{D}[\theta,{\bm u}]=\exp{\sum_{\mathbf{e}\in\Sigma}in\,\theta_\mathbf{e}(\nabla\times {\bm u})_\mathbf{e}}.\label{eq:ZD}
\eeq
Note that, the support of $\mathcal{D}$ is not $\Sigma$ alone, but $\Sigma\bigcup\tilde{\Sigma}$, where $\tilde{\Sigma}$ is the boundary of the dual lattice of $L_2$, formed by plaquettes piercing links $\left<\mathbf{ej}\right>$ with $\mathbf{e}\in\Sigma$ and $\mathbf{j}\in L_2$. See Fig.~\ref{fig:dual} for illustration.

In continuum, the duality defect is represented in terms of an integral on the surface $\Sigma$,
\beq
D[\theta,a]=\exp\Big(\frac{in}{2\pi}\int_\Sigma\theta\mathrm{d}a\Big),\label{eq:S}
\eeq
in which the flux quantization of $U(1)$ gauge field $a$ is $\int\mathrm{d}a\in 2\pi\mathbb{Z}$. It reduces to the non-invertible T-duality defect of the compact free boson in (1+1)D~\cite{Shao2024,Choi2022a,Argurio2025}.

{\it Non-invertibility of the duality defect.---}
From the half duality transformation, $\mathcal{D}$ in \Eq{eq:ZD} represents the particle--vortex duality defect. To test whether it is invertible or not, we adopt a standard approach to fuse $\mathcal{D}$ with its reverse $\bar{\mathcal{D}}$, and compute the fusion result $\mathcal{C}=\mathcal{D}\times\bar{\mathcal{D}}$~\cite{Shao2024}. Consider two adjacent surfaces $\Sigma_1$ and $\Sigma_2$ in the cubic lattice $L$ and perform duality transformation only on the links $\left<\mathbf{e}_1\mathbf{e}_2\right>$ connecting $\Sigma_1$ and $\Sigma_2$, which yields
\beq
(\mathcal{D}\times\bar{\mathcal{D}})[\theta,{\bm u}]=\sum_{\{{\bm u}\}}\exp{\sum_{\mathbf{e}_{1,2}}}in(\theta_{\mathbf{e}_1}-\theta_{\mathbf{e}_2})(\nabla\times {\bm u})_{\mathbf{e}_1\mathbf{e}_2}.
\eeq
Next, one in principle needs to identify the supports of $\mathcal{D}$ and $\bar{\mathcal{D}}$ to fuse them. However, on the lattice this is not exactly achievable, as the supports of $\mathcal{D}$ and $\bar{\mathcal{D}}$ differ by a half lattice translation. Formally, identifying their supports amounts to identify $\Sigma_1$ and $\Sigma_2$ as $\Sigma$, view $\theta_{\mathbf{e}_1}$ and $\theta_{\mathbf{e}_2}$ as field variables living on the two sides of $\Sigma$, and summing over the surface degree of freedom of the ${\bm u}$ field~\cite{Choi2022a}. Denote $\delta\theta_{\mathbf{e}}$ as the jump between field variables on site $\mathbf{e}$ but different sides of $\Sigma$. As each $u_\mathbf{rr^\prime}$ couples to a $(\delta\theta_{\mathbf{e}}-\delta\theta_{\mathbf{e^\prime}})$, the fusion result of $\mathcal{D}$ and $\bar{\mathcal{D}}$ upon integrating out $l_{\mathbf{e}_1\mathbf{e}_2}$ in the topological limit is,
\beq
\mathcal{C}[\theta]=\prod_{\left<\mathbf{ee^\prime}\right>\in\Sigma}\sum_{z_{\mathbf{e}\mathbf{e^\prime}}\in\mathbb{Z}_n}\frac{1}{n}\delta_{2\pi}\Big(\delta\theta_{\mathbf{e}}-\delta\theta_{\mathbf{e^\prime}}-\frac{2\pi}{n}z_{\mathbf{ee^\prime}}\Big),\label{eq:C}
\eeq
in which $z_{\mathbf{ee^\prime}}$ defines a dynamical $\mathbb{Z}_n$ gauge theory on the 2D surface $\Sigma$. The delta function $\delta_{2\pi}$ is periodic, which imposes flatness of the $\mathbb{Z}_n$ gauge field on each plaquette $\square\in\Sigma$, $(\nabla\times{\bm z})_\square=0$ (mod $n$), from summation around $\square$. Therefore, fusing $\mathcal{D}$ with its orientation reverse produces a surface hosting a (1+1)D $\mathbb{Z}_n$ gauge theory, which is not equivalent to identity unless $n=1$~\cite{Roumpedakis2023,Shao2024}. For $n=1$, the expression of $\mathcal{C}$, \Eq{eq:C}, imposes $\delta\theta_\mathbf{e}=\mathrm{Const.}$, which indeed glues the XY models in two sides of $\Sigma$ as a trivial surface after modding out the global $U(1)$ shift.

Typically, existence of non-invertible symmetry and defect suggests anomaly. However, non-invertibility of the particle--vortex duality defect $\mathcal{D}$ has a different origin. It stems from the non-invertibility of the duality transformation. Note that, the charge-$n$ XY model \Eq{eq:H} with $K^\prime=0$ has a built-in $\mathbb{Z}_n$ branch structure on each site: $\theta_\mathbf{i}\mapsto\theta_\mathbf{i}+2\pi q_\mathbf{i}/n$ with $q_\mathbf{i}\in\mathbb{Z}_n$. This $\mathbb{Z}_n$ branch is not merely an artificial overcounting, as $\theta_\mathbf{i}$ itself represents a physical operator in the charge-$n$ XY model. For instance, in a vestigial charge-$2ne$ superconductor, $e^{in\theta_\mathbf{i}}$ is the superconducting order parameter, while $e^{i\theta_\mathbf{i}}$ is the microscopic charge-$2e$ Cooper pair. In this case, the $\mathbb{Z}_n$ branches of $\theta_\mathbf{i}$ are distinguishable by physical electric charge probes. Therefore, sequentially performing the duality and the inverse duality shifts $\theta_\mathbf{i}$ to $\theta_\mathbf{i}+2\pi q_\mathbf{i}/n$, which does not result in an identical mapping. Instead, it lands in a different physical configuration of $\theta_\mathbf{i}$. Further inclusion of a weak $K^\prime$ term in \Eq{eq:H} with $K^\prime<n^2K$ lifts the exact degeneracy of different $\mathbb{Z}_n$ branches; however, the branch structure is not destroyed. Thus, the duality defect remains non-invertible. On the other hand, a strong $K^\prime$ term eliminates the $\mathbb{Z}_n$ branches, which effectively converts the charge-$n$ XY model to the ordinary charge-$1$ XY model and makes the duality defect invertible.

A complementary way to understand the non-invertibility is from the operator correspondence between dual theories. A local operator across an invertible duality defect should land in a local operator in the dual theory. However, across a non-invertible duality defect it becomes nonlocal~\cite{Shao2024}. This is most conveniently seen in the continuous form of the duality defect. Consider a vertex operator $V_p(\mathbf{x})=e^{ip\theta(\mathbf{x})}$ with integer $p\notin n\mathbb{Z}$ inserted in the action \Eq{eq:S}. Equation of motion of $\theta$ yields $\mathrm{d}a/(2\pi)=(p/n)\delta(\mathbf{x})$. Therefore, across the duality defect the vertex operator $V_p(\mathbf{x})$ becomes a monopole carrying fractional magnetic charge $p/n$, which is nonlocal unless $n=1$. This corroborates the duality defect being non-invertible for $n>1$.

\begin{figure}[htbp]
  \centering
  \includegraphics[width=0.66\linewidth]{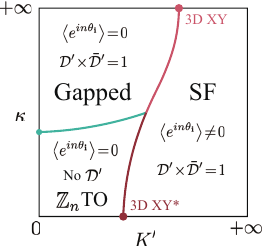}
  \caption{Schematic phase diagram of the gauged XY model \Eq{eq:Hz} as a function of $K^\prime$ and $\kappa$. The 3D XY and XY$^*$ transitions are marked in light and dark red, respectively. The confinement--deconfinement transition of $\mathbb{Z}_n$ gauge theory is marked in turquoise. In the $\mathbb{Z}_n$ TO phase and the 3D XY$^\ast$ transition line, $\mathcal{D}^\prime$ is not a standalone duality defect as it is not gauge invariant. In the boson gapped phase and the SF phase, $\mathcal{D}^\prime$ flows to an invertible duality defect. }
  \label{fig:phase}
\end{figure}

{\it Promoting $\mathbb{Z}_n$ branch to dynamical gauge field.---}
As discussed in the previous section, the charge-$n$ XY model \Eq{eq:H} has a $\mathbb{Z}_n$ branch structure, which leads to the non-invertibility of the particle--vortex duality defect. It is then natural to study the fate of the non-invertible duality defect when the $\mathbb{Z}_n$ branch structure is promoted to be dynamical. This amounts to couple a dynamical $\mathbb{Z}_n$ gauge field $z_\mathbf{ij}\in\mathbb{Z}_n$ to the $\theta_\mathbf{i}$ field,
\bea
\beta H^\prime =&-K^\prime\sum_{\left<\mathbf{ij}\right>}\cos{\Big(\theta_\mathbf{i}-\theta_\mathbf{j}-\frac{2\pi}{n}z_\mathbf{ij}\Big)}\\
&-\frac{1}{2\kappa}\sum_\square\cos{\frac{2\pi}{n}(\nabla\times {\bm z})}.\label{eq:Hz}
\eea
Here $K^\prime>0$ and $\kappa$ represent the Higgs and the Maxwell coupling, respectively. The phase diagram of this model is shown in Fig.~\ref{fig:phase}. In the gauged model, the $K\cos{(n\theta_\mathbf{i}-n\theta_\mathbf{j})}$ term does not qualitatively affect the phase diagram and hence omitted~\cite{Gao2022}. With $\kappa$ sufficiently large, the $\mathbb{Z}_n$ gauge field is confined, which binds $n$ elementary boson described by $e^{i\theta}$ to a bound state $e^{in\theta}$. The system then undergoes a 3D XY transition without fractionalization from the boson trivially gapped phase to the SF phase upon increasing $K^\prime$. For $\kappa$ sufficiently small, the $\mathbb{Z}_n$ gauge field is deconfined, which yields $\mathbb{Z}_n$ topological order (TO) phase at small $K^\prime$~\cite{Xu2012}. When $K^\prime$ is increased, condensation of $e^{i\theta}$ Higgses the $\mathbb{Z}_n$ gauge field and lands in the SF phase. This phase transition is a 3D XY$^*$ transition, as the gauge invariant order parameter remains $e^{in\theta_\mathbf{i}}$, instead of $e^{i\theta_\mathbf{i}}$~\cite{Isakov2012,Xu2012}.

Half duality transformation factorizes the partition function of Hamiltonian \Eq{eq:Hz} after Villainization into three pieces, 
\beq
Z^\prime=\int D[\theta,{\bm z},{\bm u}]\,W_1^\prime[\theta,{\bm z}]\times\widetilde{W}_2^\prime[{\bm z},{\bm u}]\times\mathcal{D}^\prime[\theta,{\bm u}],
\eeq
where $W_1^\prime$ remains the Boltzmann weight of Villainized \Eq{eq:Hz} on $L_1$. Dual theory on $L_2$ is described by $\widetilde{W}_2^\prime$,
\begin{widetext}
\beq
\widetilde{W}_2^\prime[{\bm z},{\bm u}]=\exp\bigg[\sum_\square-\frac{(\nabla\times {\bm u})^2}{2K^\prime}+\frac{1}{2\kappa}\cos{\frac{2\pi}{n}(\nabla\times {\bm z})}-\sum_{\left<\mathbf{ij}\right>}\frac{2\pi i}{n}(\nabla\times {\bm u})_\mathbf{ij}\,z_\mathbf{ij}\bigg],\label{eq:Z2p}
\eeq
\end{widetext}
where $u_\mathbf{rr^\prime}$ is an integer field defined on dual lattice links. It can be similarly converted to a noncompact $U(1)$ gauge field $a_\mathbf{rr^\prime}$ by introducing an integer vortex current field $J_\mathbf{rr^\prime}$ and performing Poisson's summation. The last piece $\mathcal{D}^\prime$ is now
\beq
\mathcal{D}^\prime[\theta,{\bm u}]=\exp{\sum_{\mathbf{e}\in\Sigma}i\theta_\mathbf{e}(\nabla\times {\bm u})_\mathbf{e}}.
\eeq
Note that $\mathcal{D}^\prime$ is no longer gauge invariant. Therefore, it should not be viewed as a genuine duality defect, but merely a duality wall. With a dynamical $\mathbb{Z}_n$ gauge field, its invertibility is not well-defined. In the IR of the $\mathbb{Z}_n$ TO phase, it remains gauge covariant, while in the IR of the boson gapped or SF phases where the $\mathbb{Z}_n$ gauge field is confined or Higgsed, it flows to an {\it invertible} gauge invariant duality defect. 

When the Maxwell coupling $\kappa$ is small, the $\mathbb{Z}_n$ gauge field is deconfined, and $z_\mathbf{ij}$ cannot be integrated out in \Eq{eq:Z2p}. Correspondingly, the duality wall is not gauge invariant, and there is no well-defined duality defect in the $\mathbb{Z}_n$ TO phase. When $\kappa$ is large, the $\mathbb{Z}_n$ gauge field is confined, and $z_\mathbf{ij}$ can be integrated out. Deep in the confined limit $\kappa\rightarrow+\infty$, explicitly summing over $z_\mathbf{ij}$ restricts the integer field $\nabla\times {\bm u}$ to be divisible by $n$. Rescaling $\nabla\times {\bm u}$ to $(\nabla\times {\bm u})/n$ and using Poisson's summation identifies $\widetilde{W}_2^\prime$ to $\widetilde{W}_2$ described by \Eq{eq:Z2} with $K=K^\prime/n^2$, and $\mathcal{D}^\prime$ to $\mathcal{D}$ described by \Eq{eq:ZD}. It seems to suggest a non-invertible duality defect. However, $e^{i\theta_\mathbf{i}}$ is not a gauge invariant operator, while the composite $e^{in\theta_\mathbf{i}}$ remains in the physical operator content. Therefore, $\mathcal{D}^\prime$ should be defined in terms of $\vartheta_\mathbf{i}=n\theta_\mathbf{i}$, which makes it an invertible duality defect of the charge-$1$ XY model in low energy. Physically, the vestigial charge-$2ne$ superconductor picture ceases to apply here, as Cooper pair is always a physical operator remaining in the physical operator content in a charge-$2ne$ superconductor.

This can also be seen from Higgsing the $\mathbb{Z}_n$ gauge field. Consider a sufficiently large $K^\prime$, where the theory remains in the SF phase with the $\mathbb{Z}_n$ gauge field $z_\mathbf{ij}$ Higgsed regardless of $\kappa$. See the phase diagram Fig.~\ref{fig:phase}. From the dual theory \Eq{eq:Z2p} perspective, field $u_\mathbf{rr^\prime}$ can be integrated out by taking $K^\prime\rightarrow+\infty$, which indeed fixes $\nabla\times {\bm z}=0$. Locally, $z_\mathbf{ij}$ becomes a pure gauge and can be gauge-fixed to $z_\mathbf{ij}=0$, while possible global sectors are identified by the condensed unit-charge matter. Physically, with large $K^\prime$, the vortices in the dual model are tightly bound, and the noncompact $U(1)$ gauge field $a_\mathbf{rr^\prime}$ converted from $u_\mathbf{rr^\prime}$ through Poisson's summation is in the Coulomb phase. The topological $U(1)$ symmetry with conserved current ${\bm j}=(\nabla\times {\bm a})/(2\pi)$ is spontaneously broken. This Higgses $z_\mathbf{ij}$ through the lattice BF coupling between $a_\mathbf{rr^\prime}$ ($u_\mathbf{rr^\prime}$) and $z_\mathbf{ij}$ in \Eq{eq:Z2p}. Deep in the IR of the Higgsed phase, gauge field $z_\mathbf{ij}$ is entirely gapped out and decoupled from the low energy sector. As a result, the duality wall $\mathcal{D}^\prime$ flows to an invertible duality defect.

In summary, the duality wall $\mathcal{D}^\prime$ flows to an invertible duality defect in the IR of the boson gapped phase and the SF phase. See Fig.~\ref{fig:phase} for illustration. This is exactly the effect of promoting the $\mathbb{Z}_n$ branch structure to a dynamical $\mathbb{Z}_n$ gauge field in the charge-$n$ XY model. Upon gauging, the (1+1)D $\mathbb{Z}_n$ gauge field that appears as independent topological sectors on the fused defect $\mathcal{C}$ described by \Eq{eq:C} becomes the connecting condition of the bulk gauge field across the interface. Since bulk gauge fields are summed over in the partition function and gauge equivalent configurations are identified, those formerly distinct defect sectors no longer label distinct physical fusion channels, which removes the non-invertibility. Pictorially, the $\mathbb{Z}_n$ gauge field living on the duality defect is absorbed and mixed into the bulk. In addition, gauging $\mathbb{Z}_n$ inevitably changes the physical local operator algebra. The $\theta_\mathbf{i}$ field is no longer a gauge invariant observable, and the order parameter is a composite, which gives rise to the 3D XY$^*$ universality class. The change in the order parameter critical exponent is the physical consequence of the same finite-sector structure that appears as the non-invertibility of the duality defect in the ungauged theory.

{\it Outlook.---}
The idea of half dualization can be applied in several directions. A first application is to the easy-plane noncompact $\mathbb{CP}^1$ model~\cite{Motrunich:2004}, which possesses self-duality as a generalization of the particle--vortex duality. However, we expect the duality defect in this model to be invertible, similar to the charge-$1$ XY model. A more promising direction is boson--fermion duality. Exact lattice constructions of three-dimensional bosonization~\cite{Chen2018,Aitken2018,Son2019} suggest that one can implement half dualization to such a lattice model and examine the duality defect. Such a duality defect is possibly related to the fermion parity symmetry and/or spin structure of the manifold, which may yield non-invertibility. Finally, it would be useful to understand half dualization directly in continuum spacetime. TQFTs provide a natural setting for this question, since different continuum Lagrangians can describe the same TQFT and interfaces between such descriptions realize duality defects.

{\it Acknowledgment.---} The author is grateful to Jichen Feng, Zhaoyu Han, Chunxiao Liu, Taige Wang, Yan-Qi Wang, and Hui Yang for providing helpful feedback on the manuscript. The author acknowledges Srinivas Raghu and Ruihua Fan for inspiring discussions in the early stage. The author is supported by the Berkeley graduate program.

\bibliographystyle{apsrev4-2}

\bibliography{bibs.bib}

\appendix
\section{End Matter}

With a nonvanishing $K^\prime$ term in \Eq{eq:H}, the Villainized partition function becomes~\cite{Gao2022},
\beq
Z=\int D[\theta]\prod_{\left<\mathbf{ij}\right>}\sum_{m^\prime_\mathbf{ij}\in\mathbb{Z}}\sum_{p_\mathbf{ij}\in\mathbb{Z}_n}V_\mathbf{ij},\label{eq:ZEM}
\eeq
where the Boltzmann weight $V_\mathbf{ij}$ is,
\beq
V_\mathbf{ij}&=&\exp{\bigg[-\frac{n^2K}{2}\Big(\theta_\mathbf{i}-\theta_\mathbf{j}-2\pi m^\prime_\mathbf{ij}-\frac{2\pi}{n}p_\mathbf{ij}\Big)^2\bigg]}\nn\\
&&\times\exp{\bigg[-\Delta\Big(1-\cos{\frac{2\pi p_\mathbf{ij}}{n}}\Big)\bigg]}.
\eeq
Here $\Delta\approx K^\prime/(1-\cos{(2\pi/n)})$. For $K^\prime>n^2K$, the charge-$n$ XY model does not possess local minima near $\theta_\mathbf{i}-\theta_j=2\pi p_\mathbf{ij}/n$ for $p_\mathbf{ij}\neq 0$~\cite{Lee:1985aa,Gao2022}. Correspondingly, the Villainized partition function is qualitatively equivalent to that of a charge-$1$ XY model with only global minima $\theta_\mathbf{i}=\theta_\mathbf{j}$. This eliminates the local $\mathbb{Z}_n$ branch structure. For $K^\prime<n^2K$, the $\mathbb{Z}_n$ branch remains.

Assume $K^\prime<n^2K$ and define $m_\mathbf{ij}=nm^\prime_\mathbf{ij}+p_\mathbf{ij}$. Then the $m_\mathbf{ij}$ and $\theta_\mathbf{ij}$ part of \Eq{eq:ZEM} recovers \Eq{eq:Z0} in the main text. The $p_\mathbf{ij}$ part is organized to an emergent $\mathbb{Z}_{n}$ gauge sector, as shown in Ref.~\cite{Gao2022}. We first derive the partition function of the $K^\prime=0$ case. From \Eq{eq:Z0}, we first implement Poisson's summation over $m_\mathbf{ij}$ for every link $\left<\mathbf{ij}\right>\in L_2$. The partition function then reads,
\beq
Z=\int D[\theta]\prod_{\left<\mathbf{ij}\right>\in L_1}\sum_{m_\mathbf{ij}}\exp\bigg[-\frac{K}{2}(n\theta_\mathbf{i}-n\theta_\mathbf{j}-2\pi m_\mathbf{ij})^2\bigg]\nn\\
\times\prod_{\left<\mathbf{ij}\right>\in L_2}\sum_{l_\mathbf{ij}}\exp\bigg[-\frac{l_\mathbf{ij}^2}{2K}+inl_\mathbf{ij}(\theta_\mathbf{i}-\theta_\mathbf{j})\bigg],\nn
\eeq
where $l_\mathbf{ij}$ is an integer variable defined on links in $L_2$. Integration over $\theta_\mathbf{i}$ with $\mathbf{i}\in L_2$ enforces $\nabla\cdot n{\bm l}_\mathbf{i}=0$, where ${\bm l}_\mathbf{i}=(l_{\mathbf{i},\mathbf{i}+\hat{x}},l_{\mathbf{i},\mathbf{i}+\hat{y}},l_{\mathbf{i},\mathbf{i}+\hat{z}})$. As $n$ is an integer, this constraint is equivalent to $\nabla\cdot {\bm l}_\mathbf{i}=0$. However, the coupling between $\theta_\mathbf{e}$ on site $\mathbf{e}\in\Sigma$ and $l_\mathbf{ej}$ remains in the partition function. After integration, the partition function becomes,
\beq
Z=\int D[\theta]\prod_{\left<\mathbf{ij}\right>\in L_1}\sum_{m_\mathbf{ij}}\exp\bigg[-\frac{K}{2}(n\theta_\mathbf{i}-n\theta_\mathbf{j}-2\pi m_\mathbf{ij})^2\bigg]\nn\\
\times\prod_{\left<\mathbf{ij}\right>\in L_2}\sum_{\nabla\cdot {\bm l}_\mathbf{i}=0}\exp\bigg(\!-\frac{l_\mathbf{ij}^2}{2K}\bigg)\prod_{\mathbf{e}\in\Sigma}\exp{(in\theta_\mathbf{e}l_{\mathbf{ej}})}.\nn
\eeq
To solve the constraint $\nabla\cdot {\bm l}_\mathbf{i}=0$, we introduce a dual integer field $u_\mathbf{rr^\prime}$ defined on the dual lattice links, satisfying $l_\mathbf{ij}=(\nabla\times {\bm u})_\mathbf{ij}$, where the curl of $u_\mathbf{rr^\prime}$ is defined on the dual lattice plaquette that pierces the lattice link $\left<\mathbf{ij}\right>$. On torus, $l_\mathbf{ij}=(\nabla\times {\bm u})_\mathbf{ij}$ should be supplemented by integer flux sectors that can be suppressed via summation over twisted boundary conditions ($U(1)$ orbifolding). With $u_\mathbf{rr^\prime}$, the partition function reads,
\beq
Z=\int D[\theta]\prod_{\left<\mathbf{ij}\right>\in L_1}\sum_{m_\mathbf{ij}}\exp\bigg[-\frac{K}{2}(n\theta_\mathbf{i}-n\theta_\mathbf{j}-2\pi m_\mathbf{ij})^2\bigg]\nn\\
\times\!\!\prod_{\left<\mathbf{ij}\right>\in L_2}\sum_{u_\mathbf{rr^\prime}}\exp\bigg[-\frac{(\nabla\times {\bm u})^2_\mathbf{ij}}{2K}\bigg]\exp{\sum_{\mathbf{e}\in\Sigma}in\theta_\mathbf{e}(\nabla\times {\bm u})_\mathbf{ej}},\nn
\eeq
which recovers \Eq{eq:Zf} in the main text. By introducing an integer vortex current $J_\mathbf{rr^\prime}$ satisfying $\nabla\cdot {\bm J}_\mathbf{r}$ and using Poisson's summation, $u_\mathbf{rr^\prime}$ is converted to a noncompact $U(1)$ gauge field $a_\mathbf{rr^\prime}$~\cite{Gao2022}. The Boltzmann weight of the dual theory becomes
\beq
\widetilde{W}_2[{\bm a}]=\sum_{\{\nabla\cdot{\bm J}_\mathbf{r}=0\}}\exp\bigg[\sum_\mathbf{r}-\frac{(\nabla\times {\bm a})^2}{8\pi^2K}+i{\bm a}_\mathbf{r}\cdot {\bm J}_\mathbf{r}\bigg],\label{eq:Z2l}
\eeq
and the duality defect becomes
\beq
\mathcal{D}[\theta,{\bm a}]=\exp{\sum_{\mathbf{e}\in\Sigma}\frac{in}{2\pi}\,\theta_\mathbf{e}(\nabla\times {\bm a})_\mathbf{e}},\label{eq:Da}
\eeq
which is corresponding to the continuum form \Eq{eq:S} in the main text. 

With inclusion of a finite $K^\prime<n^2K$, the duality defect remains \Eq{eq:Da}, while the dual theory is supplemented by a $\mathbb{Z}_n$ gauge theory piece as shown in Ref.~\cite{Gao2022},
\beq
\widetilde{W}_2[{\bm a}]=\sum_{\{\nabla\cdot{\bm J}_\mathbf{r}=0\}}\exp\bigg[\sum_\mathbf{r}-\frac{(\nabla\times {\bm a})^2}{8\pi^2K}+i{\bm a}_\mathbf{r}\cdot {\bm J}_\mathbf{r}\bigg]\nn\\
\times \sum_{\{\sigma_\mathbf{rr^\prime}\}}\left<\exp{\Big(\frac{2\pi i}{n}}{\bm \sigma}_\mathbf{r}\cdot {\bm J}_\mathbf{r}\Big)\right>_{\mathbb{Z}_n}Z_{\mathbb{Z}_n}.
\eeq
Here $Z_{\mathbb{Z}_n}$ is the partition function of a $\mathbb{Z}_n$ gauge theory defined in dual link variables $\sigma_\mathbf{rr^\prime}\in\mathbb{Z}_n$, and $\left<\cdots\right>_{\mathbb{Z}_n}$ is the expectation value of the Wilson loop under the $\mathbb{Z}_n$ gauge theory. The $\mathbb{Z}_n$ gauge theory arising from finite $K^\prime$ only modifies the bulk vortex current weights but the duality defect $\mathcal{D}[\theta,{\bm a}]$. Therefore the fractional-monopole diagnostic remains valid. This can be clearly seen from the continuum. In continuum, the dual theory is described by the Lagrangian,
\beq
\mathcal{L}=|D_{a+\alpha}\varphi|^2+t|\varphi|^2+s|\varphi|^4+\frac{in}{2\pi}\alpha\mathrm{d}\tilde{\alpha}+\frac{1}{2g}(\mathrm{d}\alpha)^2,
\eeq
where $\varphi$ is the vortex field whose current is corresponding to $J_\mathbf{rr^\prime}$, mutual Chern--Simons term between $\alpha$ and $\tilde{\alpha}$ suggests $\alpha$ being a $\mathbb{Z}_n$ gauge field, and $g$ is the Maxwell coupling of $\alpha$. In this field theory, as $\varphi$ carries charge $1$ under $a$, the flux of $a$ is quantized by $2\pi$. Therefore, similar to that in the main text, a vertex operator $V_p=e^{ip\theta(\mathbf{x})}$ in the charge-$n$ XY model across the duality defect \Eq{eq:S} becomes a monopole carrying fractional magnetic charge $p/n$, which is nonlocal unless $n=1$. This indicates the non-invertibility of the duality defect when the charge-$n$ XY model possesses a finite $K^\prime<n^2K$ term.

\end{document}